\documentclass[12pt]{article}
\usepackage{graphicx}
\usepackage{graphics}
\title{On certain sums over the nontrivial zeta zeros} 
\author{Mark W. Coffey\\
Department of Physics\\
Colorado School of Mines\\
Golden, CO  80401\\
(Received $\mbox{~~~~~~~~~~~~~~~~~~~~~~~~~~~~~~~2009}$)}
\date{October 2, 2009}
\begin{document}
\maketitle
\baselineskip=25 pt
\begin{abstract}

We study coefficients $b_n$ that are expressible as sums over the Li/Keiper
constants $\lambda_j$.  We present a number of relations for and
representations of $b_n$.  These include the expression of $b_n$ as a sum 
over nontrivial zeros of the Riemann zeta function, as well as integral
representations.  Conditional on the Riemann hypothesis, we provide the
asymptotic form of $b_n \sim 2^{-n-2}\ln n$.

\end{abstract}
 
\vspace{.25cm}
\baselineskip=15pt
\centerline{\bf Key words and phrases}
\medskip 

\noindent

Riemann zeta function, Li criterion, Li/Keiper constants, Riemann hypothesis, functional equation, nontrivial zeros, integral representation, asymptotic form 

\vfill
\centerline{\bf MSC numbers}
11M06, 11M26, 11Y60 
 
\baselineskip=25pt
\pagebreak
\medskip
\centerline{\bf Introduction} 
\medskip

Let $\zeta$ denote the Riemann zeta function, and $\xi(s)=(s/2)(s-1)\pi^{-s/2}
\Gamma(s/2)\zeta(s)$ the classical completed zeta function, where $\Gamma$ is
the Gamma function \cite{edwards,ivic,riemann,titch}.  
Within the critical strip ${\cal S}$, $0 < \mbox{Re} ~s <1$, the complex zeros of $\zeta$ and $\xi$ coincide, and we denote them by $\rho$.  The $\xi$-function is 
entire, of order $1$, and of maximal type.

Herein, we mainly investigate certain sums $b_n$ over complex zeta function zeros.
We provide various representations and properties of these sums.
We also supply some remarks on the Li criterion \cite{li} for the Riemann hypothesis (RH).  


We recall that the Li equivalence for the RH results 
as a necessary and sufficient condition that the logarithmic derivative 
of the function $\xi[1/(1-z)]$ be analytic in the unit disk.  This obtains
from a conformal map of the critical strip to this disk.  The equivalence 
\cite{li} states that a necessary and sufficient condition for the nontrivial 
zeros of the $\zeta$-function to lie on the critical line Re $s=1/2$ is that 
constants $\{\lambda_k\}_{k=1}^\infty$ are nonnegative for every integer $k$.
The sequence $\{\lambda_n\}_{n=1}^\infty$ can be defined by
$$\lambda_n = {1 \over {(n-1)!}}{d^n \over {ds^n}}[s^{n-1}\ln \xi(s)]_{s=1}.
\eqno(1.1)$$
The $\lambda_j$'s are connected to sums over the nontrivial zeros of $\zeta(s)$ 
by way of \cite{keiper,li}
$$\lambda_n=\sum_\rho\left [1-\left(1-{1 \over \rho}\right)^n \right ].
\eqno(1.2)$$     
For further discussion of the Li criterion, its application, and results on
series expansion of the $\xi$ function, see for instance 
\cite{bombieri,coffey04,coffey03,coffeyijcms,coffeyprsa,coffeyrmjm}.     

In particular, consider the expansions \cite{heetal}
$$\ln\varphi(z) \equiv \ln \xi\left({1 \over {1-z}}\right)=-\ln 2+\sum_{n=1}^\infty {\lambda_n \over n} z^n=b_0+\sum_{n=1}^\infty b_n(z+1)^n.  \eqno(1.3)$$
The middle expansion in terms of the Li/Keiper constants $\lambda_n$ holds 
for $|z|<\delta_1 <1$, where $\delta_1 =1$
corresponds to the Riemann hypothesis.  Similarly, the right-most expansion
holds for $|z+1|<\delta_2<2$.  Thus, (1.3) has overlapping domains of
expansion, allowing analytic continuation.  

In \cite{heetal}, the coefficients $b_n$ of (1.3) are simply an inessential
device.  However, we treat their properties, including a ``curious identity"
$b_2=b_3$, in the next section.  The latter relation is simply the beginning
of an infinite set of relations that we make explicit.
In the Appendix, we record approximate numerical values for the early
coefficients.  Reference \cite{heetal} contains physical discussion,
including attempting to regard $\xi$ as a quantum mechanical wave function.

To emphasize that the $b_n$'s are not required for the purposes of 
\cite{heetal}, we have the following argument.  Let $N(T)$ be the count of
nontrivial zeta zeros for $0<\mbox{Im} ~\rho<T$.  We have
$N(T)=\pi^{-1} \mbox{Im} \ln \xi(1/2+i T)$, and it is well known that as
$T \to \infty$, $N(T)=(T/2\pi)\ln(T/2\pi)-T/(2\pi)+O(\ln T)$.  
Suppose that the RH holds.  Then we have
$$\lambda_n=2\sum_{j=1}^\infty (1- \cos n \theta_j) \geq 0, \eqno(1.4)$$
where $\rho=1/2\pm i \mu_j$ or $\mu_j=(1/2)\cot(\theta_j/2)$.  We can
rewrite (1.4) as
$$\lambda_n=2\int_0^\infty [1-\cos\theta(\mu)]dN(\mu),  \eqno(1.5)$$
where the lower limit just as well may be taken as $\mu_1$.  
Integrating by parts, we have
$$\lambda_n=-2n\int_0^\infty \sin n\theta {{d\theta} \over {d\mu}}N(\mu)
d\mu+2(1-\cos n\theta)N(\mu)|_0^\infty,  \eqno(1.6)$$
where $\mu(\theta)={1 \over 2}\cot(\theta/2)$.
For $\mu \to \infty$, we have $\theta=1/\mu+O(1/\mu^3)$,
$1-\cos n\theta=-n^2/2\mu^2+O(1/\mu^3)$, and the required limit on the
right side of (1.6) is zero.  We therefore obtain the equivalent exact
forms
$$\lambda_n=-2n\int_0^\infty \sin n\theta {{d\theta} \over {d\mu}}N(\mu)
d\mu  \eqno(1.7)$$
$$=2n\int_0^{\pi/2} \sin n\theta N(\mu) d\theta(\mu).  \eqno(1.8)$$
We can note that in (1.8), $\pi/2$ may be replaced by $2\cot^{-1}(2\mu_1)
\simeq 1/\sqrt{2}10$, and therefore the range of integration is a relatively
narrow one.  

We may quickly rewrite (1.7), as $d\theta/d\mu=-4/(4\mu^2+1)$.
Furthermore, $\sin n\theta=\sin \theta U_{n-1}(\cos \theta)$, where
$U_k$ is the $k$th Chebyshev polynomial of the second kind, and 
$\cos \theta=(4\mu^2-1)/(4\mu^2+1)$.  We obtain
$$\lambda_n=32n\int_0^\infty {{\mu N(\mu)} \over {(4\mu^2+1)^2}} U_{n-1}
\left({{4\mu^2-1} \over {4\mu^2+1}}\right)d\mu, \eqno(1.9)$$
thereby recovering (3.13) of \cite{heetal}.  As observed there, on the
RH, the values $\lambda_n$ are indeed nonnegative.  

Implicit in \cite{heetal} (p. 8) are the relations
$$-\ln 2=b_0+\sum_{n=1}^\infty b_n, \eqno(1.10)$$  
and for $n \geq 1$,
$$b_n=\sum_{j=n}^\infty (-1)^{j-n} {j \choose n}{\lambda_j \over j}. 
\eqno(1.11)$$
That (1.11) holds may be easily verified by using \cite{heetal} (3.4) and the orthogonality relation
$$\sum_{m=n}^j (-1)^m {m \choose n}{j \choose m}=(-1)^j \delta_{jn}, \eqno(1.12)$$
with $\delta_{jk}$ the Kronecker symbol.  We shall have recourse to these 
relations in the following developments.

\medskip
\centerline{\bf Relations and representations of $b_n$}
\medskip

We have
{\newline \bf Proposition 1}.  For $n \geq 1$ we have
$$b_n={1 \over {n 2^n}}\sum_{\rho \in {\cal S}}\left[1-\left(1+{1 \over {1-2\rho}}
\right)^n\right].  \eqno(2.1)$$

{\bf Corollary 1}.  We have $b_1=0$.  

{\bf Corollary 2}.  We have $\zeta'(1/2)/\zeta(1/2)={1 \over 2}(\gamma+{\pi \over
2}+3\ln 2+\ln \pi)$.

In (2.1) the sum includes zeros $\rho$ along with $1-\rho$.  (Owing to the functional equation of the $\xi$ function or $\zeta$-functions.)  We write
$\sum_\rho$ when the companion zero $1-\rho$ is explicitly taken into account.

We have
{\newline \bf Corollary 3}.
$$b_n=-{1 \over {n2^n}}\sum_{j=1}^n {n \choose j}\sum_{\rho \in {\cal S}}{1 \over {(1-2\rho)^j}}.
\eqno(2.2)$$
Thus
$$b_n=-{1 \over {n2^{n-1}}}\sum_{k=2}^{[n/2]} {n \choose {2k}}\sum_\rho {1 \over {(1-2\rho)^{2k}}}.  \eqno(2.3)$$

We have
\newline{\bf Proposition 2}.  The coefficient $b_{2n+1}$ is always expressible 
as a rational linear combination of $b_{2n}$, $b_{2n-2}$, ..., $b_2$.

{\bf Examples}.  
We have $b_3=b_2$, $b_5=2b_4-b_3$, and $b_7=3b_3-5b_4+3b_6$.

Let $\Sigma_{2k} \equiv \sum_{\rho \in {\cal S}} {1 \over {(1-2\rho)^{2k}}}$.  
We have
{\newline \bf Corollary 4}.  We have the relation for $n$ even
$$b_{n+1}={n \over 2}b_n+{1 \over {(n+1)}}{1 \over 2^n}\sum_{k=1}^{n/2-1}\left[
n{n \choose {2k}}-{n \choose {2k-1}}\right] \Sigma_{2k}.  \eqno(2.4)$$

We have
\newline{\bf Proposition 3}.  We have the summation relation for $n \geq 2$,
$$[1-(-1)^n]b_n=\sum_{j=2}^{n-1} (-1)^j {{n-1} \choose {j-1}} b_j.  \eqno(2.5)$$

{\bf Corollary 5}.  In particular, we have $b_2=b_3$.

Let $L_n^\alpha$ be the Laguerre polynomial of degree $n$ and parameter $\alpha$
\cite{andrews}.  Then we have the following representation.

{\bf Proposition 4}.  We have
$$b_n={1 \over {n2^{n+1}}}\sum_{\rho \in {\cal S}}\int_0^\infty e^{-\rho u} 
L_{n-1}^1\left({u \over 2}\right)du.  \eqno(2.6)$$

{\bf Corollary 6}.  On the RH, with $\rho=1/2+it_j$, and $t_j$ is real, we have
$$b_n={1 \over {n2^{n}}}\sum_{j=1}^\infty \int_0^\infty e^{-u/2} \cos(t_j u)
L_{n-1}^1\left({u \over 2}\right)du.  \eqno(2.7)$$


Write \cite{li}
$$\varphi(z)=1+\sum_{j=1}^\infty a_j z^j,  \eqno(2.8)$$
with $\xi(1/2)=1+\sum_{j=1}^\infty (-1)^ja_j=\exp(b_0)$ on the RH.  
The rapid asymptotic growth of $a_j$ with $j$ has been described in 
\cite{coffeyjat}.  We have
{\newline \bf Proposition 5}.  We have the recurrence relation for $m \geq 1$
$$\xi(1/2)(m+1)b_m=\sum_{j=m}^\infty (-1)^{j-m} (j+1){j \choose m}a_{j+1}
-\sum_{\ell=1}^m \sum_{j=\ell}^\infty (-1)^{j-\ell}{j \choose \ell} (m-\ell+1)b_{m-\ell}a_j .  \eqno(2.9)$$

{\bf Proposition 6}.  On the RH, we have for $n \geq 1$,
$$b_n=2^{-n}\int_0^{\pi/2} {{\sin[(n-1)\theta/2]} \over {\cos^{n+1}(\theta/2)}}
N(\mu)d\theta.  \eqno(2.10)$$

{\bf Corollary 7}.  On the RH, we have for $n \gg 1$,
$$b_n \sim 2^{-n-2}\left[\ln\left(n-1\right)+\gamma-1-\ln(4\pi)\right]. \eqno(2.11)$$

{\bf Proposition 7}.  
On the RH, we have
$$\lambda_n={n \over 2}\ln n+ (\gamma-1-\ln 2\pi)n + o(n).  \eqno(2.12)$$


A Corollary of Proposition 7 is Corollary 7.

In the next section, proofs are supplied, as well as some discussion.

\medskip
\centerline{\bf Proofs of Propositions}
\medskip

{\it Proposition 1}.  We substitute the sum (1.2) into (1.11), 
$$b_n=\sum_{\rho \in {\cal S}} \sum_{j=n}^\infty {{(-1)^{j-n}} \over j}
{j \choose n} \left[1-\left(1-{1 \over \rho}\right)^j\right]$$
$$={1 \over {n 2^n}}\sum_{\rho \in {\cal S}}\left[1-\left(1+{1 \over {1-2\rho}}
\right)^n\right].  \eqno(3.1)$$
The interchange of sums is justified by the absolute convergence of (1.2).

Corollary 1 immediately follows as we have
$$b_1=-{1 \over 2}\sum_\rho\left({1 \over {1-2\rho}}-{1 \over {1-2\rho}}\right)=0.  \eqno(3.2)$$

From the Hadamard product for the $\xi$-function, we have
$${{\xi'(z)} \over {\xi(z)}}={1 \over z}\sum_{\rho \in {\cal S}} {1 \over {1-
\rho/z}}.  \eqno(3.3)$$
Therefore, $b_1=-{1 \over 4}{{\xi'} \over \xi}\left({1 \over 2}\right)$,
implying Corollary 2.  

Corollary 3 (2.2) follows by binomial expansion in (2.1).  

{\it Remarks}.  Corollary 2 recovers what otherwise may be found by applying the
functional equation of the $\zeta$ function.  

Indeed, all odd order derivatives of $\xi$ are zero at $1/2$.  

{\it Proposition 2}.  We use (2.3) and put $\Sigma_{2k} \equiv \sum_{\rho \in
{\cal S}} {1 \over {(1-2\rho)^{2k}}}$.
Then, for each $n$, $\Sigma_{2n}$ may be eliminated between $b_{2n}$ and $b_{2n+1}$, and the result follows.

{\it Corollary 4}.   We have for $n$ even from Corollary 3
$$b_n=-{1 \over {n2^{n-1}}}\sum_{k=1}^{n/2} {n \choose {2k}}\Sigma_{2k},
\eqno(3.4a)$$
and
$$b_{n+1}=-{1 \over {(n+1)2^n}}\sum_{k=1}^{n/2} {{n+1} \choose {2k}}\Sigma_{2k}.
\eqno(3.4b)$$
Therefore, from (3.4a) we have
$$\Sigma_n=-n2^{n-1}b_n-\sum_{k=1}^{n/2-1}{n \choose {2k}}\Sigma_{2k}.  
\eqno(3.5)$$
We insert this equation into (3.4b) written in the form
$$b_{n+1}=-{1 \over {(n+1)2^n}}\left[\sum_{k=1}^{n/2-1} {{n+1} \choose {2k}}\Sigma_{2k}+(n+1)\Sigma_n\right].  \eqno(3.6)$$
We find
$$b_{n+1}={n \over 2}b_n-{1 \over {(n+1)}}{1 \over 2^n}\sum_{k=1}^{n/2-1}
\left[{{(n+1)} \choose {2k}}-(n+1){n \choose {2k}}\right]\Sigma_{2k}.  
\eqno(3.7)$$
Finally, we use a recursion relation for the binomial coefficient,
${{m+1} \choose n}={m \choose n}+{m \choose {n-1}}$, and obtain (2.4).

{\it Examples}.  We have $\Sigma_4=8(-4b_4+3b_3)$, $b_5=2b_4-b_3$, $\Sigma_6=-12(25b_3-40b_4+16b_6)$, and $b_7=(1/56)(-57b_3+80b_4+24b_6)$.  Of course, $b_3=b_2=-(1/4)\Sigma_2$.

{\it Proposition 3}.  The result is a consequence of the functional equation 
of the $\xi$ function, so that
$$\xi \left({1 \over {1-z}}\right)=\xi\left({z \over {z-1}}\right)=\varphi(z).
\eqno(3.8)$$
We have 
$$\ln\xi\left({z \over {z-1}}\right)=b_0+\sum_{n=1}^\infty b_n \left(1+{1 \over z}\right )^n$$
$$=b_0+\sum_{n=1}^\infty b_n (1+z)^n \sum_{k=n-1}^\infty (-1)^n {k \choose {n-1}}
(1+z)^{k-n+1}, \eqno(3.9)$$
using 
$${1 \over q^j}=\sum_{k=j-1}^\infty {k \choose {j-1}} (1+q)^{k-j+1}.  \eqno(3.10)$$
Then
$$\ln\xi\left({z \over {z-1}}\right)=b_0+\sum_{n=1}^\infty b_n\sum_{k=n}^\infty
(-1)^n {{k-1} \choose {n-1}}(1+z)^k$$
$$=b_0+\sum_{k=1}^\infty \sum_{n=1}^k b_n(-1)^n{{k-1} \choose {n-1}}(1+z)^k.
\eqno(3.11)$$
Comparing with the expansion (1.3), we obtain
$$b_n=\sum_{j=2}^n (-1)^j {{n-1} \choose {j-1}}b_j, \eqno(3.12)$$
where we have used $b_1=0$.  Relation (2.5) follows.

{\it Proposition 4}.  We use (1.11) and \cite{coffeyjat} (Prop. 1)
$$\lambda_j=\sum_{\rho \in {\cal S}}\int_0^\infty e^{-\rho u}L_{j-1}^1(u)du,
\eqno(3.13)$$
so that
$$b_n=\sum_{\rho \in {\cal S}}\int_0^\infty e^{-\rho u}(-1)^n\sum_{j=n}^\infty {{(-1)^j} \over j}{j \choose n}L_{j-1}^1(u) du.  \eqno(3.14)$$
We use the representation \cite{andrews} (p. 286) with Bessel function $J_\alpha$ 
for $\alpha>-1$,
$$L_n^\alpha(x)={{e^x x^{-\alpha/2}} \over {n!}}\int_0^\infty t^{n+\alpha/2} J_\alpha (2\sqrt{xt})e^{-t} dt, \eqno(3.15)$$
giving
$$\sum_{j=n}^\infty {{(-1)^j} \over j}{j \choose n}L_{j-1}^1(u)
=e^u u^{-1/2} \sum_{j=n}^\infty {{(-1)^j} \over {j!}}{j \choose n}\int_0^\infty
t^{j-1/2}J_1(2\sqrt{ut})dt$$
$$=e^u u^{-1/2}{{(-1)^n} \over {n!}}\int_0^\infty e^{-2t}t^{n-1/2}J_1(2\sqrt{ut})
dt$$
$$=2e^u u^{-1/2}{{(-1)^n} \over {n!}}\int_0^\infty e^{-2x^2} x^{2n} J_1(2\sqrt{u}
x)dx.  \eqno(3.16)$$
The integral is first evaluated \cite{grad} (p. 716) in terms of the
confluent hypergeometric function $_1F_1$:
$$\sum_{j=n}^\infty {{(-1)^j} \over j}{j \choose n}L_{j-1}^1(u)
={{(-1)^n} \over 2^{n+1}} e^u ~_1F_1\left(n+1;2;-{u \over 2}\right)$$
$$={{(-1)^n} \over 2^{n+1}} ~_1F_1\left(1-n;2;{u \over 2}\right)$$
$$={{(-1)^n} \over 2^{n+1}}{1 \over n}L_{n-1}^1\left({u \over 2}\right).
\eqno(3.17)$$
Here we have used Kummer's first transformation for the function $_1F_1$
\cite{andrews} (p. 191) as well as the relation
$$L_n^\alpha(z)={{n+\alpha} \choose n} ~_1F_1(-n,\alpha+1;z).  \eqno(3.18)$$  
The insertion of (3.17) into (3.14) gives the Proposition.

{\it Remarks}.  By integrating by parts we may verify that (2.6) returns
the sum representation of Proposition 1.  We have
$$b_n={1 \over {n 2^n}}\sum_{\rho \in {\cal S}} \int_0^\infty e^{-2\rho v}
L_{n-1}^1(v)dv$$
$$=-{1 \over {n 2^n}}\sum_{\rho \in {\cal S}} \int_0^\infty e^{-2\rho v}
{d \over {dv}}L_n(v)dv.  \eqno(3.19)$$
The integral converges since necessarily Re $\rho>0$.  We use the Laplace
transform of a Laguerre polynomial \cite{grad} (p. 844), and we recover (2.1).

The theory of Laguerre polynomials is pervasive in formulating the Li criterion
\cite{coffey03,coffeyjat,coffeyrmjm}.

By multiply differentiating (3.17), we have a family of summations,
$$\sum_{j=n}^\infty {{(-1)^j} \over j}{j \choose n}L_{j-k-1}^{k+1}(u)
={{(-1)^n} \over 2^{n+k+1}}{1 \over n}L_{n-k-1}^{k+1}\left({u \over 2}\right).
\eqno(3.20)$$
Otherwise, we may follow the steps as above and find for $\alpha>-1$ and
$z \neq 1$
$$\sum_{j=n}^\infty {{z^j} \over j}{j \choose n}L_{j-1}^\alpha(u)
={z^n \over {(1-z)^{n+\alpha}}}{1 \over n}L_{n-1}^\alpha\left({u \over {1-z}}
\right).  \eqno(3.21)$$

We have the contour integral representation
$$L_n^\alpha(z)={{\Gamma(n+\alpha+1)} \over {2\pi i n!}}\int_{-\infty}^{(0+)} 
\left(1-{z \over t}\right)^n e^t {{dt} \over t^{\alpha+1}}, \eqno(3.22)$$
where the contour encircles the origin in the positive direction and closes at
Re $z=-\infty$.  
This gives
$$\sum_{j=n}^\infty {{(-1)^j} \over j}{j \choose n}L_{j-1}^1(u)
={{(-1)^n} \over {2\pi i}} {1 \over 2^{n+1}} \int_{-\infty}^{(0+)} 
{{(1-u/t)^{n-1}} \over {(1-u/2t)^{n+1}}}e^t {{dt} \over t^2}.  \eqno(3.23)$$
It may be verified that the residue at $t=u/2$ gives $L_{n-1}^1(u/2)/n$.

The defining sums of Proposition 1 may be recovered from Corollary 6 in the
following way.  Write on the RH
$$b_n=-{1 \over {n2^{n-1}}}\sum_{j=1}^\infty \int_0^\infty e^{-v} \cos(2t_j v) 
{d \over {dv}} L_n(v)dv$$
$$={1 \over {n2^{n-1}}}\sum_{j=1}^\infty \left\{\int_0^\infty \left[{d \over {dv}}e^{-v}\cos(2t_jv)\right]L_n(v)dv + 1\right\}.  \eqno(3.24)$$ 
Then use \cite{grad} (p. 846) for $n$ even and odd to evaluate the integrals.

{\it Proposition 5}.  This follows from the identity ${{\varphi'} \over \varphi}
\varphi=\varphi'$.  We make use of
$$1+\sum_{j=1}^\infty a_jz_j=1+\sum_{j=1}^\infty a_j\sum_{\ell=0}^j {j \choose 
\ell} (-1)^{j-\ell} (z+1)^\ell$$
$$=1+\sum_{j=1}^\infty (-1)^j a_j + \sum_{\ell=1}^\infty \sum_{j=\ell}^\infty
(-1)^{j-\ell}{j \choose \ell}a_j (z+1)^\ell, \eqno(3.25)$$
and similarly 
$$\varphi'(z)=\sum_{j=1}^\infty ja_j z^{j-1}=\sum_{\ell=0}^\infty 
\sum_{j=\ell}^\infty (-1)^{j-\ell} (j+1){j \choose \ell}a_{j+1}(z+1)^\ell .
\eqno(3.26)$$
With some further series manipulations we obtain (2.9).

{\it Proposition 6}.  We have from (1.11) and (1.8)
$$b_n=2\sum_{j=n}^\infty (-1)^{j-n}{j \choose n}\int_0^{\pi/2} \sin j\theta
N(\mu)d\theta.  \eqno(3.27)$$
We then apply the sum $\sum_{j=n}^\infty (-1)^j {j \choose n}x^j=(-1)^n x^n
(1+x)^{-n-1}$ at $x=\exp(\pm i\theta)$.  Then simple manipulations yield
(2.10).

{\it Corollary 7}.  In (2.10) we change variable to $x=(n-1)\theta/2$.  At the
leading order in $n$ we obtain
$$b_n \sim 2^{-n-1} {1 \over \pi}\int_0^\infty {{\sin x} \over x} \left[\ln\left({{n-1} \over {4\pi x}}\right) -1\right]dx.  \eqno(3.28)$$
Performing the integral gives (2.11).

{\it Remarks}.  It is evident that (2.10) includes the cases $b_1=0$,
$b_2=b_3$, and $b_5=2b_4-b_3$.  

The asymptotic result (2.11) is consistent with the right-most expansion in (1.3)
having a radius of convergence at most $2$.

Simply from the sum representation (2.1) one may suspect an asymptotic form
$b_n \sim {1 \over 2^n}{1 \over \mu_1}$.  One could also estimate $b_n$ from
$$b_n={1 \over {2\pi i}}\int_C {{\varphi(z)dz} \over {(z+1)^{n+1}}}, \eqno(3.29)$$
where the contour encircles $z=-1$.

For one of the integrals in (3.28), we first have
$\int_0^\infty x^\alpha \sin x ~dx=\cos(\pi \alpha/2)\Gamma(\alpha+1)$ for
$-2<\mbox{Re} ~\alpha<0$.  Then performing logarithmic differentiation and taking
$\alpha\to -1$ we obtain
$$\int_0^\infty {{\sin x} \over x}\ln \left({1 \over x}\right) dx={\pi \over 2}\gamma.  \eqno(3.30)$$

Numerically from (2.1) as a sum over the first $10^5$ nontrivial zeta zeros
\cite{odlyzko} we find $b_{1000} \simeq 9.21 \times 10^{-302}$ while from (2.11)
we have $b_{1000} \simeq 9.22 \times 10^{-302}$.  See figure 1 for a semilog
plot of the first $1000$ values of $b_n$ versus $n$ (with $b_1$ omitted).
These numerical values, suggesting that indeed the right-most expansion in (1.3) has radius of convergence $2$, could be taken as evidence that the RH holds.

{\it Proposition 7}.  Method 1.  Similarly to Corollary 7, we have using (1.8),
$$\lambda_n=2\int_0^{n/\sqrt{2}10} \sin x ~N\left[\mu\left({x \over n}\right)
\right] dx.  \eqno(3.31)$$
Therefore, at the leading order we have
$$\lambda_n \sim {n \over \pi}\int_0^\infty {{\sin x} \over x}\left[\ln\left(
{n \over {2\pi x}}\right)-1\right]dx,  \eqno(3.32)$$
with the error incurred being $o(n)$.  Using (3.30) gives the Proposition.

Method 2.  We alternatively use the expression (1.9) and have the expansions
$$\theta(\mu)={1 \over \mu}-{1 \over {12\mu^3}}+O\left({1 \over \mu^5}\right),
\eqno(3.33)$$
and $\sin(n\theta)=\sin(n/\mu)+O(1/\mu^3)$.  We have
$$U_{n-1}(\cos \theta)={{\sin(n\theta)} \over {\sin \theta}}={{(4\mu^2+1)} \over
{4\mu}} \sin(n\theta)=\left[\mu+O\left({1 \over \mu}\right)\right]\sin(n\theta).
\eqno(3.34)$$
We then have
$$\lambda_n \sim 32n\int_{\mu_1}^\infty {{\mu^2 N(\mu)} \over {(4\mu^2+1)^2}}
\sin\left({n \over \mu}\right) d\mu \sim 2n\int_{\mu_1}^\infty {{N(\mu)} \over
\mu^2} \sin\left({n \over \mu}\right)d\mu$$
$$=2 \int_0^{n/\mu_1} N\left({n \over v}\right) \sin v dv$$
$$\sim -{n \over \pi}\int_0^\infty {{\sin v} \over v}\left[\ln\left({{2\pi v}
\over n}\right)+1\right]dv.  \eqno(3.35)$$
Using (3.30) we find (2.12).

{\it Corollary 7}.  We now reprove this Corollary as a result of Proposition 7.
We have by (1.11),
$$b_n=\sum_{j=n}^\infty (-1)^{j-n} {j \choose n}\left[{1 \over 2}\ln j+\gamma-1
-\ln(2\pi)+o(1)\right].  \eqno(3.36)$$
In order to accurately approximate the summand, we use the digamma function
$\psi=\Gamma'/\Gamma$, and have $\psi(j)=\ln j +o(1)$ for $j \gg 1$.  Then we have,
using an integral representation for $\psi$ \cite{grad} (p. 943), 
$$b_n=\sum_{j=n}^\infty (-1)^{j-n}{j \choose n}\left\{{1 \over 2}[\psi(j)+\gamma]
+{\gamma \over 2}-1-\ln(2\pi)+o(1)\right\}$$
$$=\sum_{j=n}^\infty (-1)^{j-n}{j \choose n}\left[{1 \over 2}\int_0^1 \left( {{t^{j-1} -1} \over {t-1}}\right) dt + {\gamma \over 2}-1-\ln(2\pi)+o(1)\right]$$
$$={1 \over 2}\int_0^1 \left[{t^{n-1} \over {(t+1)^{n+1}}}-{1 \over 2^{n+1}}
\right]{{dt} \over {(t-1)}}+\left[{\gamma \over 2}-1-\ln(2\pi)\right]{1 \over 2^{n+1}}+{{o(1)} \over 2^{n+1}}$$
$$={1 \over 2}{1 \over 2^{n+1}}\left[\psi(n)+\gamma-\ln 2 -{1 \over n}\right]
+[{\gamma \over 2}-1-\ln(2\pi)]{1 \over 2^{n+1}} +{{o(1)} \over 2^{n+1}}$$
$$={1 \over 2^{n+2}}\ln n + +\left(\gamma-1-\ln\pi-{3 \over 2}\ln 2\right)
{1 \over 2^{n+1}} + {{o(1)} \over 2^{n+1}}.  \eqno(3.37)$$

{\it Remarks}.
The result (2.12) is not new \cite{coffey03}, but we include it and the method
of proof as a companion to Proposition 6 and Corollary 7.  We suspect that the $o(n)$ terms in (2.12) are of size $O(n^{1/2+\epsilon})$ for any $\epsilon>0$.

Generally alternating binomial sums may be difficult to estimate, but we have
done so in recovering Corollary 7.

Regarding (3.36) and (3.37), it is possible to use an even more accurate 
approximation to $\ln j$, with $\psi(j+1/2)=\ln j +O(1/j^2)$, but at the cost 
of a more complicated integral to perform.

Suppose that $\lambda_j$ has a subdominant term close to $\sqrt{j}$.  Then we
expect there to be a correction term in $b_n$ close to
$$\sum_{j=n}^\infty (-1)^{j-n} {j \choose n}{1 \over j^{1/2}} \sim
\sum_{j=n}^\infty (-1)^{j-n} {j \choose n}{{\Gamma(j)} \over {\Gamma(j+1/2)}}$$
$$={{\Gamma(n)} \over {\Gamma(n+1/2)}} ~_2F_1\left(n,n+1;n+{1 \over 2};-1\right)
\sim {\sqrt{2} \over 2^{n+1}}{1 \over \sqrt{n}}.  \eqno(3.38)$$
Here, $_2F_1$ is the Gauss hypergeometric function \cite{andrews,grad}, and by
transformation rules \cite{grad} (p. 1043), the $_2F_1$ function in (3.35) is
the same as
$$2^{-n-1} ~_2F_1\left(n,{1 \over 2};n+{1 \over 2};{1 \over 2}\right)
=2^{-n-1}\sqrt{2} ~_2F_1\left(-{1 \over 2},{1 \over 2};n+{1 \over 2};-1\right).
\eqno(3.39)$$

The above argument extends so that if $\lambda_j$ has a subdominant term
$j^{1/2+\epsilon}$, we expect in $b_n$ a term close to
$$\sum_{j=n}^\infty (-1)^{j-n} {j \choose n}{{\Gamma(j+\epsilon)} \over {\Gamma(j+1/2)}}
={{\Gamma(n+\epsilon)} \over {\Gamma(n+1/2)}} ~_2F_1\left(n+\epsilon,n+1;n+{1 \over 2};-1\right).  \eqno(3.40)$$

\pagebreak
\centerline{\bf Appendix:  Values of $b_n$}

Exact expressions for $b_n$ can be written in terms of $\ln \pi$, polygammic
constants $\psi^{(j)}(1/4)$, and the derivatives $\zeta^{(k)}(1/2)$.  The
following table gives approximate numerical values for the initial $b_n$'s.

\begin{table}[h]\centering
\begin{tabular}{||r|r||} \hline
$n$ & $b_n$ \\
\hline
0    & -0.698922\\
\hline
1    & 0 \\  
\hline
2    & 0.00144406 \\
\hline
3    & 0.00144406 \\
\hline
4    &  0.00108297  \\ \hline
5    & 0.000721886 \\  
\hline
6    & 0.000451088 \\
\hline
7    & 0.00027058 \\
\hline
8    &  0.000157786  \\ \hline 
9    & 0.0000901269 \\  
\hline
10    & 0.0000506726 \\
\hline
11    & 0.0000281364 \\
\hline
12    & 0.0000154657   \\ \hline
13    & $8.43018 \times 10^{-6}$ \\  
\hline
14    & $4.56299 \times 10^{-6}$ \\
\hline
15    & $2.45502 \times 10^{-6}$\\
\hline
16    &  $1.31 \times 10^{-6}$  \\  \hline 
17    &  $6.99 \times 10^{-7}$  \\  \hline 
18    &  $3.71 \times 10^{-7}$  \\  \hline
19    &  $1.96 \times 10^{-7}$  \\  \hline 
20    &  $1.03 \times 10^{-7}$  \\  \hline  
21    &  $5.44 \times 10^{-8}$  \\  \hline
22    &  $2.85 \times 10^{-8}$  \\  \hline
23    &  $1.49 \times 10^{-8}$  \\  \hline
24    &  $7.79 \times 10^{-9}$  \\  \hline
25    &  $4.06 \times 10^{-9}$  \\  \hline             
\end{tabular}
\end{table}

The values $b_0,\ldots,b_{15}$ have been obtained in Mathematica by series
expansion of the $\varphi$ function, (1.3).  The remaining values have been
found in Matlab by summing over the first $10^5$ complex zeta zeros 
\cite{odlyzko}, (2.1).

\begin{figure}[hb]
\begin{center}
\includegraphics[height=4.0in,width=5.5in,angle=0]{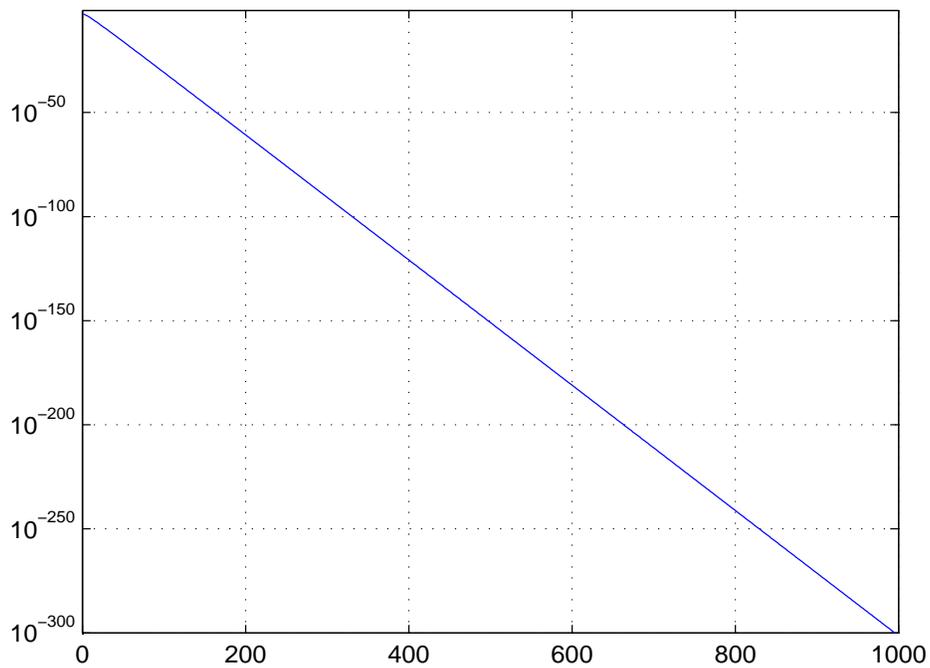}
\caption{A semilog plot of the first $1000$ values of $b_n$ obtained from
(2.1) by summing over the first $10^5$ complex zeta zeros \cite{odlyzko}.
The value $b_1=0$ is omitted.}
\end{center}
\end{figure}

  
\pagebreak

\end{document}